\documentclass[aps,prl,twocolumn,preprintnumbers,nofootinbib]{revtex4}
\usepackage{graphicx}
\usepackage{amssymb}
\usepackage{amsmath,mathrsfs,verbatim}
\usepackage{times}
\usepackage{latexsym}
\def\beq{\begin{equation}}
\def\eeq{\end{equation}}
\def\bey{\begin{eqnarray}}
\def\eey{\end{eqnarray}}

\def\lsim{\mathrel{\raise.3ex\hbox{$<$\kern-.75em\lower1ex\hbox{$\sim$}}}}
\def\gsim{\mathrel{\raise.3ex\hbox{$>$\kern-.75em\lower1ex\hbox{$\sim$}}}}

\newcommand{\be}{\begin{equation}}
\newcommand{\ee}{\end{equation}}

\newcommand{\cmsqpg}{\ {\rm cm}^2/{\rm g}}
\begin{document}

\title{Tying Dark Matter to Baryons with Self-interactions}
\author{Manoj Kaplinghat$^a$, Ryan E. Keeley$^a$, Tim Linden$^{b,c}$, and Hai-Bo Yu$^{d}$}
\affiliation{$^a$ Department of Physics and Astronomy, University of California, Irvine, California 92697, USA \\ $^b$ Department of Physics, University of California, Santa Cruz, Santa Cruz, CA, 95064, USA\\$^{c}$The Kavli Institute for Cosmological Physics, University of Chicago, Chicago, IL 60637, USA\\$^{d}$Department of Physics and Astronomy, University of California, Riverside, CA, 92507, USA}

\date{\today}

\begin{abstract}
Self-interacting dark matter (SIDM) models have been proposed to solve the small-scale issues with the collisionless cold dark matter (CDM) paradigm. We derive equilibrium solutions in these SIDM models for the dark matter halo density profile including the gravitational potential of both baryons and dark matter. Self-interactions drive dark matter to be isothermal and this ties the core sizes and shapes of dark matter halos to the spatial distribution of the stars,  a radical departure from previous expectations and from CDM predictions. Compared to predictions of SIDM-only simulations, the core sizes are smaller and the core densities are higher, with the largest effects in baryon-dominated galaxies. As an example, we find a core size around 0.3 kpc for dark matter in the Milky Way, more than an order of magnitude smaller than the core size from SIDM-only simulations, which has important implications for indirect searches of SIDM candidates.
\end{abstract}

\maketitle

{\em I. Introduction:} The CDM paradigm has been extremely successful in explaining the large-scale structure of the Universe. However, there is no established CDM-based solution to explain the central dark matter densities in galaxies. Observed dwarf galaxies and low surface brightness galaxies prefer 0.5-5 kpc cores of constant dark matter density~\cite{Simon:2004sr,KuziodeNaray:2007qi,Oh:2010ea,Walker:2011zu}, in contrast to the $1/r$ (where $r$ is distance from the center of the galaxy) cusps seen in CDM-only simulations~\cite{Navarro:1996gj}. There is also evidence for deviations from the $1/r$ behavior within the brightest (central) cluster galaxies~\cite{Newman:2012nw}. Additionally, the most massive subhalos predicted by CDM-only simulations are too dense to host observed dwarf satellite galaxies in the Milky Way~\cite{BoylanKolchin:2011dk}. It is possible that in-situ supernova feedback~\cite{Governato:2012fa}, environmental effects~\cite{Brooks:2012vi,Arraki:2012bu} or an early episode of star formation~\cite{Gritschneder:2013ai,Milosavljevic:2013eha,Amorisco:2013uwa} may play a role in resolving these issues. Here, we focus on the possibility that the above small-scale issues may be resolved by significant self-interactions among dark matter particles~\cite{Spergel:1999mh}.

Recent N-body simulations have shown that strong dark matter self-interactions can lower the central dark matter density and lead to core formation matching observations on small-scales~\cite{Rocha:2012jg,Peter:2012jh,Vogelsberger:2012ku,Zavala:2012us}. On larger scales (beyond the core), self-interacting dark matter (SIDM) behaves as the same as CDM. In particular, $\Lambda$SIDM retains all the cosmological successes of $\Lambda$CDM. However, the particle physics of SIDM models is strikingly different. For example, the existence of a $\sim$1-100 MeV light force carrier is necessary to generate the required self-scattering cross section~\cite{Tulin:2013teo}. When the mediator couples to standard model particles, it may generate signals that can be probed by direct and indirect dark matter detection experiments~\cite{Kaplinghat:2013yxa}. To quantify indirect detection signals, it is crucial to understand the SIDM halo profile in the Milky Way and its satellites. 

Elastic interactions between dark matter particles allow for energy exchange and hence transport of heat. By the time each particle has had a few interactions over the lifetime of the galaxy, an isothermal core forms~\cite{Rocha:2012jg}. Our main point in this {\em Letter} is that the presence of baryons can have a dramatic influence on the predictions for the SIDM halo profile when baryons dominate the potential well. In particular, we show that the core properties are tied to the stellar gravitational potential leading to a smaller and denser core. A straightforward conclusion from this finding is that the constraints on the self-interaction cross section will be loosened. 

To contrast our results with the expectations from SIDM-only simulations, we consider the example of the Milky Way. In the case of the dark matter dominated halos, the temperature (velocity dispersion) increases with radius in the inner region, $r\lesssim r_s$ (where the density profile is less steep than $1/r^2$). Hence interactions that lead to energy exchange between dark matter particles tend to make the inner region hotter, producing a constant-density isothermal core. The core radius is set by the transfer cross section over dark matter particle mass $\sigma_T/m_\chi$. The larger this quantity, the bigger the core with the caveat that the isothermal region is at $r\lesssim r_s$, which is true for interesting values of $\sigma_T/m_\chi$~\cite{Rocha:2012jg}. The prediction for the Milky Way core radius (where the density is half the central density) is ${\cal O}(10 {\rm kpc})$ for $\sigma_T/m_\chi \sim 1 \cmsqpg$~\cite{Rocha:2012jg,Zavala:2012us}.

If baryons dominate the potential well, as in the case of the Milky Way, they will dictate the temperature (velocity dispersion) profile of dark matter. As we will see, the dark matter temperature peaks around 1 kpc in the Milky-Way case leading to a small (sub-kpc) core size. Seen from the point of view of an equilibrium solution, the dark matter spatial density profile has to track the gravitational potential of the baryons. 
{\em Hence, we arrive at the surprising conclusion that in the limit of significant self-interactions, the radius of the dark matter core is intimately tied to the gravitational potential of the baryons.} The corresponding central density will naturally be larger than the predictions of the SIDM-only simulations. 

\begin{figure*}[t]
\includegraphics[trim=0cm 0cm 0cm 0cm, clip=true,width=0.45\textwidth]{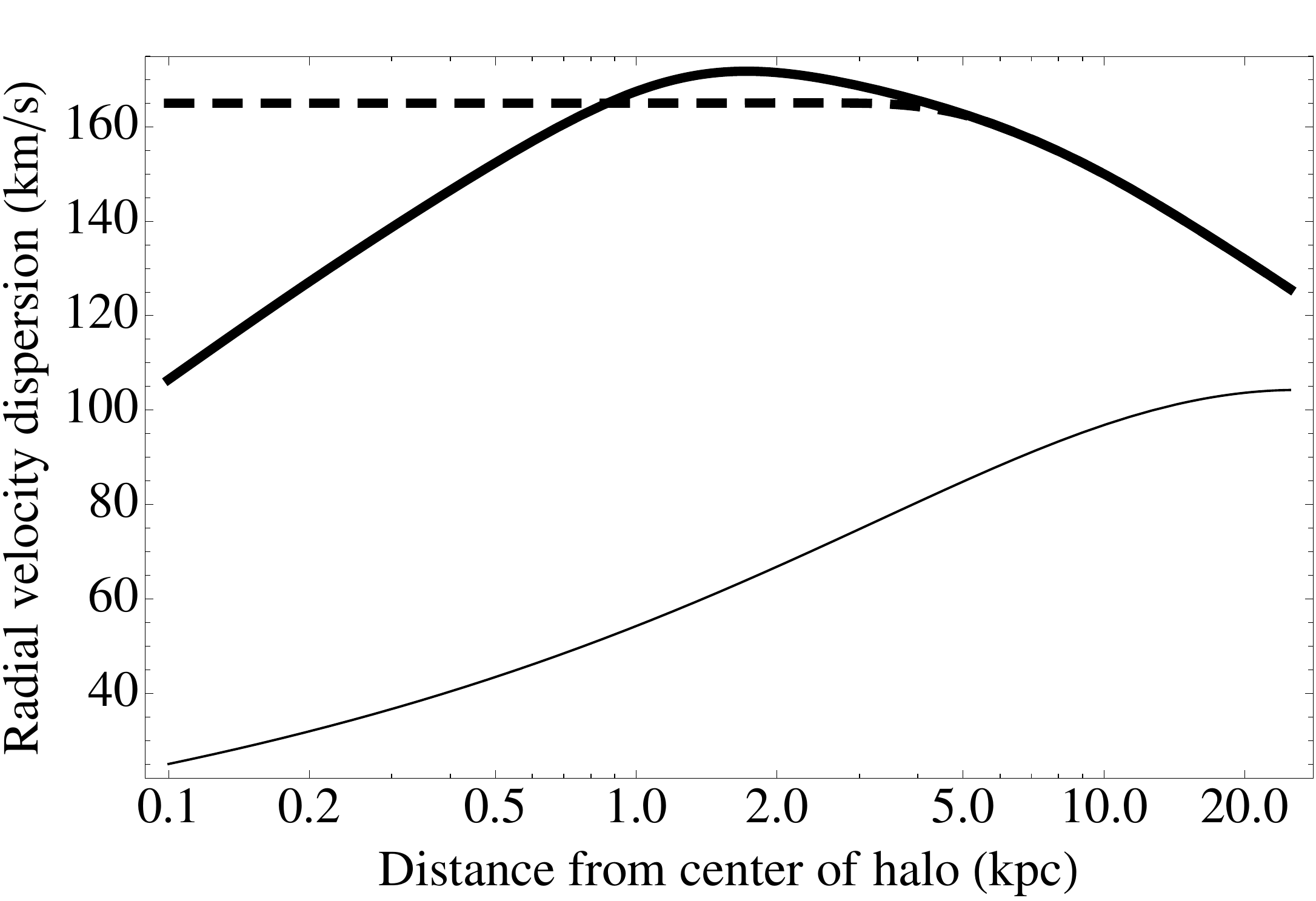}\hfill
\includegraphics[trim=0cm 0cm 0cm 0cm, clip=true,width=0.47\textwidth]{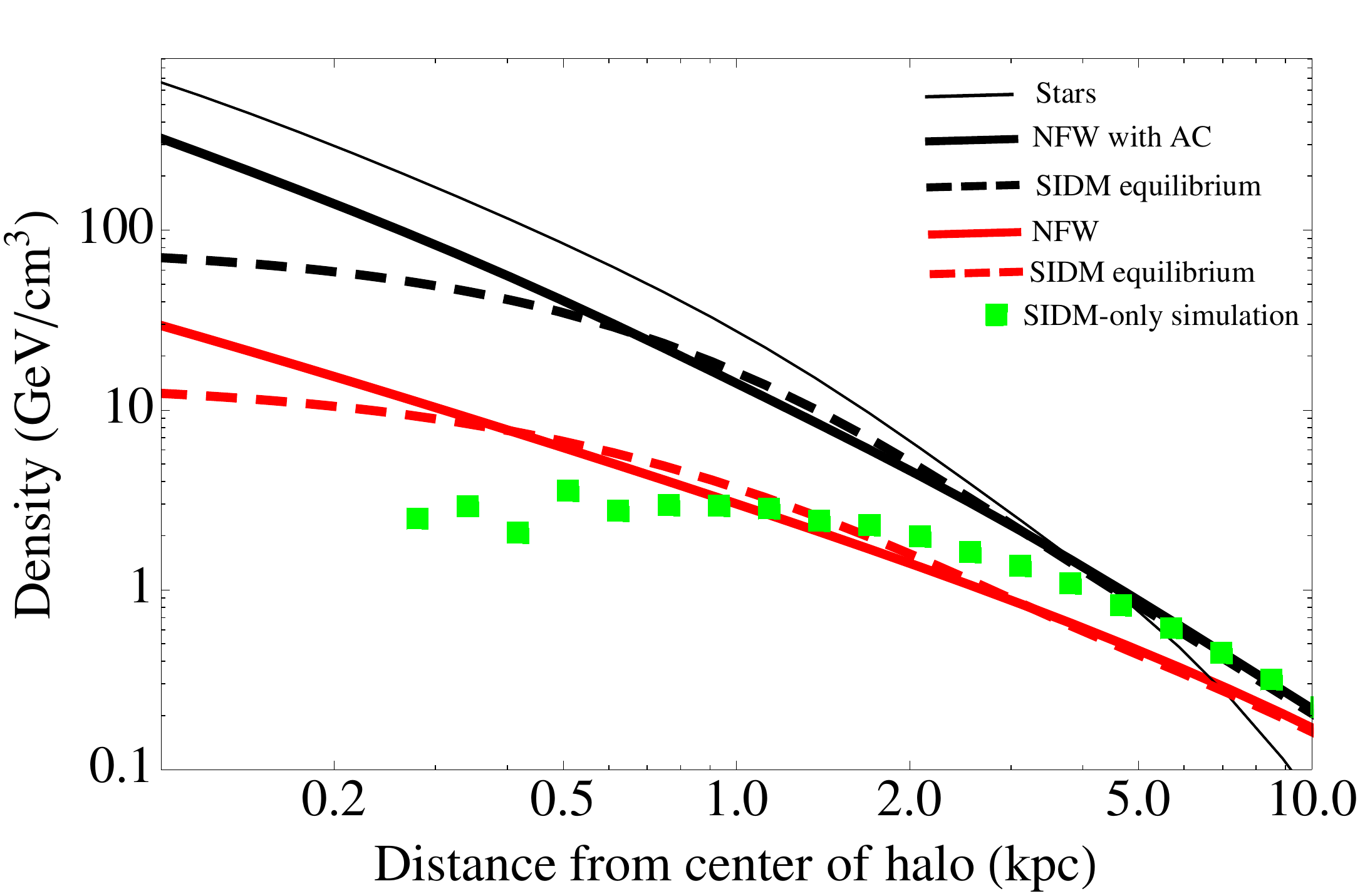}\hfill
\caption{\label{fig:density} {\it Left:} the radial velocity dispersion for dark matter (thin solid: contracted NFW; solid: contracted NFW in stellar potential; dashed: SIDM in stellar potential). {\it Right:} the dashed curves show SIDM equilibrium solutions assuming the density profile matches on to a NFW profile (solid red) and an adiabatically contracted NFW profile (solid black) at 10 kpc, and an isotropic velocity dispersion tensor. The green squares show the SIDM density profile for a $10^{12}{\rm M}_\odot$ halo from dark-matter-only simulations~\cite{Rocha:2012jg}. Note that the points below 1 kpc are not fully resolved in this simulation.}
\end{figure*}

{\em II. Solutions to the Jeans Equation:} Once an isothermal core forms, further scattering will not lead to significant changes in the density profile of the core. We neglect the possibility of core collapse here, which hasn't been seen in recent simulations with $\sigma_T/m_\chi \lesssim 1 \cmsqpg$~\cite{Rocha:2012jg,Vogelsberger:2012ku}. We further assume that the stellar profile is set on time scales shorter than the time required for dark matter to attain equilibrium through self-scattering. In this limit, we can neglect the scattering term and rewrite the Jeans equation~\cite{BT} (using Poisson's equation) with constant velocity dispersion $\sigma_0$ and dark matter density $\rho(\vec{r})=\rho_0 \exp[h(\vec{r})]$ as:
\begin{eqnarray}
\nabla_x^2 h(\vec{x}) + (4 \pi G_N r_0^2/\sigma_0^2)\left\{\rho_B(\vec{x})+\rho_0\exp[h(\vec{x})]\right\}=0\, ,\label{eq:h+phi-general}
\end{eqnarray}
where $\rho_0$ is a density scale that we take to be the central dark matter density, $\vec{x}=\vec{r}/r_0$ with $r_0$ as a length scale, and $\rho_B$ is the baryonic density profile. To illustrate our main point, let's first consider the case in which baryons completely dominate the potential well. In this limit, we can neglect the $\exp(h)$ term and the solution to Equation~\ref{eq:h+phi-general} is simply:
\begin{eqnarray}
\rho(\vec{x})=\rho_0\exp\left\{[\Phi_B(0)-\Phi_B(\vec{x})]/\sigma_0^2\right\}\,,
\label{eq:h}
\end{eqnarray}
where $\Phi_B(\vec{x})$ is the baryonic potential generated by the density distribution $\rho_B(\vec{x})$. We define the core radius as the position where the density falls by a factor of 2 or $h(\vec{r}_c)=-\ln2$. Thus, $\vec{r}_c$ is given by the solution to $\left[\Phi_B(0)-\Phi_B(\vec{r}_c)\right]=-\sigma^2_0\ln2$. It is clear that the core radius in this limit depends on the baryonic potential rather than the self-interactions as long as the interaction strength is large enough. The density profiles in generic solutions to Equation~\ref{eq:h} are also not spherically symmetric. Both these features are in marked contrast to the predictions of SIDM when dark matter dominates \cite{Rocha:2012jg,Zavala:2012us}.

To estimate the SIDM core size in the Milky Way, we specialize to the spherically symmetric solution. 
We include contributions from the stellar bulge, the thin disk and the thick disk of the Milky Way from the best-fit model advocated in Ref.~\cite{McMillan:2011wd} and then calculate the mass enclosed within spherical shells to get a ``spherical Milky Way" model. This profile turns out to be fit by a Hernquist density profile $\rho_B(r)=\rho_{B0}r_0^4/[r(r+r_0)^3]$, where we have set $r_0$ to be the Hernquist scale radius. The assumed Hernquist profile for the baryon distribution in the Milky Way can be specified by either $\Phi_B(0)=-2\pi G\rho_{B0} r^2_0 =-G M_B/r_0$ or the circular velocity $V_B^2(r_0)=-\Phi_B(0)/4$, where $G$ is  Newton's constant and $M_B$ is total mass in baryons. With $\sqrt{-\Phi_B(0)}=365~{\rm km/s}$ and $r_0=2.7~{\rm kpc}$, we found a good fit. Thus, the core radius is $r_c\approx\sigma^2_0 \ln2/(2\pi G\rho_{B0} r_0)=r_0\sigma^2_0 \ln 2/4 V_B^2(r_0) = r_0^2 \sigma^2_0 \ln2/G M_B$. Numerically, we have
\begin{eqnarray}
r_c\approx0.3~{\rm kpc}\left(\frac{r_0}{2.7~{\rm kpc}}\right)\left(\frac{\sigma_0}{150~{\rm km/s}}\right)^2\left(\frac{183~{\rm km/s}}{V_B(r_0)}\right)^2,
\end{eqnarray}
where we take a typical value $\sigma_0\sim150~{\rm km/s}$ (as we will discuss later). Thus, the expected core size in the Milky Way halo is much smaller than $\sim10$ kpc as predicted by the SIDM-only simulations.

In this spherically symmetric limit, the analytical solution for $h(x)$ can be generalized to include the case when the dark matter component is important. Assuming the above Hernquist profile for baryons, we obtain 
\begin{equation}\label{eq:h-hernquist}
\frac{1}{y^2}\frac{d}{dy}\left[y^2\frac{d}{dy}h(y)\right]+\frac{2 a_1}{y}+\frac{a_0}{(1-y)^4}\exp\left[h(y)\right]=0\,,
\end{equation}
where we define $a_0\equiv4 \pi G \rho_0 r_0^2/\sigma_0^2$ and  $a_1\equiv- \Phi_B(0)/\sigma_0^2$, and $h$ should be interpreted as a function of a new variable $y\equiv r/(r+r_0)=x/(1+x)$. The boundary conditions to solve this equation are $h(0)=0$ and $h'(0)=-a_1$, where the second term enforces a core in the center. This may be derived by noting that as $y\rightarrow 0$, the solution to $h(y)$ has to be given by $[y^2 h'(y)]'+2a_1 y =0$.

Within the core region, the density profile varies slowly. This suggests that the equation can be solved through a series of approximations. The first is obtained by setting the third term in Equation~\ref{eq:h-hernquist} equal to $a_0$ (i.e., setting $y=0$), and we get $h \approx -a_1 y -{a_0}y^2/6. $
The core radius derived from this approximate solution is given by,
\begin{equation}\label{eq:coreradius}
r_{\rm c} \approx r_0 \frac{\sqrt{1+(2/3) \ln(2) a_0/a_1^2}-1} {1+a_0/(3a_1)-\sqrt{1+(2/3) \ln(2)a_0/a_1^2}}\,.
\end{equation}
This approximation is good to about 10\% for the interesting ranges of $a_0$ and $a_1$.  
We note that if the stellar density profile differs from Hernquist and $\rho_B(r)\propto 1/r^\alpha$ for small $r$, then $h_0(y)=-2a_1y^{2-\alpha}/(2-\alpha)(3-\alpha)-a_0 y^2/6$. In particular, there is no cored profile when $\rho_B(r)$ diverges towards the center as $1/r^2$. 

Several limits of these equations are particularly illuminating. In the limit that $a_0$ is ${\cal O}(1)$ and $a_1$ is large, we obtain $r_c \approx r_0\ln(2)/[a_1-\ln(2)]$, {\em i.e.}, the core is set just by the baryonic potential, which agrees with the result we derived before. In the opposite limit when the baryons are not dynamically important, we have a self-gravitating isothermal sphere and $r_c \approx r_0 \sqrt{6 \ln(2)/a_0}$ or $r_c^2 \approx 3 \ln(2)\sigma_0^2/(2 \pi G \rho_0)$. Thus, as the baryonic contribution gets larger, the core radius becomes smaller. 

To further illuminate this result, we need estimates for the central density $\rho_0$ to fix $a_0$. 
In Ref.~\cite{Rocha:2012jg}, a model was presented for the SIDM density profile of field halos based on the radius $r_1$ where the average dark matter particle has had one interaction and the density profile in the absence of self-interactions, {\em i.e.}, the CDM halo density profile. For the Milky Way halo, the predicted CDM halo density profile has the Navarro-Frenk-White (NFW) form and we assume this profile with the appropriate concentration for  a virial mass of $10^{12}~{\rm M}_\odot$~\cite{Prada:2011jf}.  The NFW profiles have a density at the solar position $r=8.5~{\rm kpc}$ of $0.2~{\rm GeV}/{\rm cm}^3$, which is in the range of the measured value \cite{Zhang:2012rsb}. If we use a velocity dispersion of 150 km/s (appropriate for the Milky Way) the average number of scatterings per particle per 10 Gyr within the solar radius is unity for $\sigma_T/m_\chi\sim1~{\rm barn}/{\rm GeV}=0.56~{\rm cm^2/g}$. (Basically, 
$0.2~{\rm GeV}/{\rm cm}^3\ \times$ $150~{\rm km}/{\rm s}\ \times$ $1~{\rm barn}/{\rm GeV}\ \times$ $10~{\rm Gyr} \simeq 1$.)  
Hence we expect to see deviations due to self-interactions at radii smaller than $\sim$10 kpc for $\sigma_T/m_\chi\sim1~{\rm barn}/{\rm GeV}$. This cross section is consistent with all observations and is in the range required to solve the small-scale anomalies~\cite{Rocha:2012jg,Peter:2012jh}. 

Values for $\rho_0$ about 10 times the local density would be expected in SIDM simulations that do not include baryons. If this were true even when including a stellar component, then we would have $a_0$ of order unity. Specifically,
\begin{eqnarray}
a_0 \approx1\left(\frac{\rho_0}{2.2~{\rm GeV/cm}^{3}}\right) \left(\frac{r_0}{2.7~{\rm kpc}}\right)^2 \left(\frac{150~{\rm km/s}}{\sigma_0}\right)^2.
\end{eqnarray}

For the Milky Way, we will find values of $a_0={\cal O}(10)$ because the equilibrium solution including the stellar potential demands larger values of the central density $\rho_0$. 
In order to choose from the family of solutions parameterized by $a_0$ and $a_1$, we impose two conditions - that the mass within $r_1$ and the total energy within $r_1$ are the same as the halo would have had in the absence of self-interactions. These conditions are based on the model presented in Ref.~\cite{Rocha:2012jg}. 

The two resulting SIDM profiles (shown in Figure~\ref{fig:density}) show a spread of almost an order of magnitude in the the central (core) density and show that the SIDM profile depends on the details of the disk and bulge formation and associated feedback. However, the core radius is determined to be close to 0.3 kpc in both cases. We caution that this estimate depends sensitively on the assumed inner density profile of the baryons. For example, if the baryons are more centrally concentrated within 0.3 kpc, the core radius would be smaller.

For the adiabatically contracted NFW profile~\cite{Gnedin:2004cx}, the velocity dispersion profile is plotted in the right panel of Figure~\ref{fig:density}. This profile is a solution to the Jeans equation assuming that the velocity dispersion tensor is isotropic. The value of the central density $\rho_0$ is $80~{\rm GeV}/{\rm cm}^3$ and the central radial dispersion $\sigma_0$ is 165 km/s, both of which lead to an enclosed mass and energy within $r_1=15~{\rm kpc}$ (for $\sigma_T/m_\chi=1~{\rm barn}/{\rm GeV}$) equal to that of an adiabatically contracted NFW. We note that even with this high central density the mean free path is larger than the core radii.

In order to solve for the SIDM profile beyond the core region, we join the constant dispersion region smoothly to the dispersion profile in the absence of self-interactions as shown by the dashed curve in the right panel of Figure \ref{fig:density}. The density profile (dashed curves in the left panel of Figure \ref{fig:density}) is the solution to the isotropic Jeans equation assuming this velocity dispersion profile. We note that the solution in the core (for this more complete solution) is the same as Equation \ref{eq:h+phi-general}. As an aside, we find that a numerical approximation to the full range $r < r_1$ can be obtained by considering a solution of the form $h(y,p)=-a_1 y -pa_0y^2/6$ and then fixing $p$ so that the mass enclosed within $r_1$ is the same as in the case without self-interactions. 

For the NFW profile without adiabatic contraction (for example, due to significant feedback from star formation) the dispersion profile is very similar (since it is controlled in the inner regions by baryons) but the density profile is very different. We again use the isotropic Jeans equation to find the SIDM solution. Note that the assumption of an isotropic velocity dispersion tensor plays a central role in both cases. Physically, this is reasonable because scatterings lead to energy exchange that should erase the anisotropy in the velocity dispersion (which is small even in the absence of self-interactions). For this second case without initial adibatic contraction, the solution that matches the mass and energy profile in the absence of scatterings at $r_1$ has $\sigma_0=165~{\rm km}/{\rm s}$ and $\rho_0=14~{\rm GeV}/{\rm cm}^3$. 

\begin{figure}[h]
\includegraphics[trim=0cm 0cm 0cm 0cm, clip=true,width=0.4\textwidth]{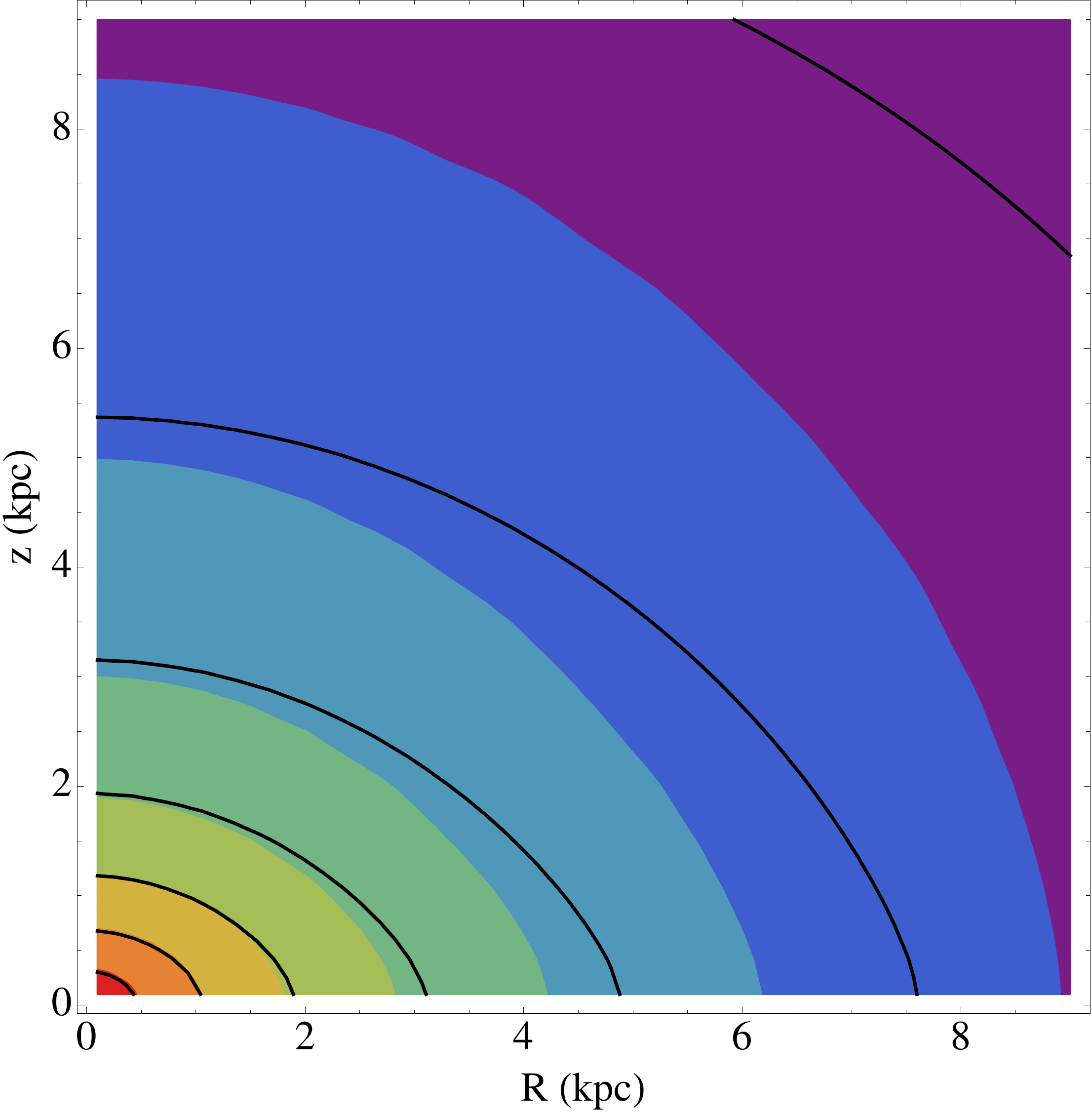}
\caption{\label{fig:shape}Constant density contours for dark matter in cylindrical coordinates ($R$, $z$) showing deviations from spherical symmetry outside the core ($\sim 0.3~{\rm kpc}$). The density at the outermost contour is $0.5~{\rm GeV}/{\rm cm}^3$ and increases by factors 2. The color shaded contours are from the full numerical analysis, while the black curves are for the approximate solution given in Equation~\ref{eq:h}.}
\end{figure}

We emphasize that the velocity dispersion profile is crucial to the effect we are pointing out. 
A core should form in the region that gets hotter (higher kinetic energy particles have larger apocenters on average) and hence the cross-over between the two dispersion profiles in Figure~\ref{fig:density} provides an estimate of the core radius of the SIDM density profile.

{\em III. Shapes of halos:} 
The shape of the halo is not expected to be spherical and deviations from spherical symmetry will depend on the stellar potential well. This has been noted previously in the context of a non-spherical isothermal solution for a halo with an embedded thin disk~\cite{2010A&A...519A..47A}. To investigate this quantitatively, we incorporate a more realistic model for the baryon distribution in the Milky Way in our analysis. In the baryon-dominated central region, we expect the simple solution given in Equation~\ref{eq:h} is valid. However, for the region away from the center, dark matter becomes important and a full numerical approach to the Jeans equation is necessary. Here, we numerically solve Equation~\ref{eq:h+phi-general} via a relaxation algorithm by rewriting Equation~\ref{eq:h+phi-general}  in the following manner, 
\newcommand{\Rt}{\tilde R}
\newcommand{\zt}{\tilde z}
\begin{equation}
\nabla^2h + a_0(\rho_B/\rho_0 + e^h)
= \partial h/\partial t\,,
\end{equation}
and demanding that $\partial h/\partial t\rightarrow 0$ for large $t$ (where $t$ plays the role of ``time" for the relaxation method). We assume axisymmetry and then approximate the above equation using finite differencing on a logarithmically spaced grid. This finite difference equation is then relaxed to an equilibrium solution in two stages (coarse and then finer spatial grid), starting with a spherical NFW density profile and the boundary conditions imposed at 10 kpc (in $R,z$) in the form of the same profile.

The SIDM constant density contours in $R,z$ for the resulting solution are plotted in Figure~\ref{fig:shape}, which clearly shows deviations from spherical symmetry when baryons dominate the potential well. We see that the approximate solution in Equation~\ref{eq:h} and the full numerical calculation give the same result in the inner region. Thus, it confirms the expectation that the SIDM distribution traces baryons when they dominate the potential well. The contours become spherical further away due to the boundary condition. Further investigations of how the shape depends on the iso-potential contours and changes away from the baryon-dominated regions (without the assumption of spherically symmetric boundary conditions) may reveal a way to use this effect to test SIDM models in galaxies and clusters.

{\em IV. Discussion:}
A natural application of the effect described above is the SIDM density profile in the centers of clusters of galaxies. Assuming a Hernquist profile and the stellar mass and effective radii in Ref.~\cite{Newman:2012nv}, we find that the core sizes are ${\cal O}(10~{\rm kpc})$ using Equation~\ref{eq:coreradius}. This is encouraging and deserves further work, especially since we predict a correlation between the SIDM core size and the effective stellar radius for which there seems to be some support~\cite{Newman:2012nw}. 

Constraints on the self-interaction strength from the observed densities and shapes in clusters of galaxies~\cite{Peter:2012jh} and the Bullet Cluster~\cite{Randall:2007ph} should be reevaluated in light of the above results.

While this effect is relevant for most galaxies, it would be particularly interesting to apply the model presented here to spiral galaxies and dwarf galaxies, which show distinct correlations in their halo core properties~\cite{2012MNRAS.420.2034S}.    

{\em V. Conclusions:} We have shown that baryons and dark matter are tied together dynamically due to self-interactions in the dark matter. The presence of baryons changes the predicted SIDM density profile by decreasing the core radius and increasing the core density, with dramatic effects in baryon-dominated galaxies. For the Milky Way halo, SIDM follows the stellar distribution and forms a core around 0.3 kpc, in contrast to the $\sim 10$ kpc core predicted in SIDM-only simulations for $\sigma_T/m_\chi\sim1~{\rm barn}/{\rm GeV}$. If SIDM is a thermal relic, the signal strength from SIDM annihilation or decay in the Galactic Center is not suppressed as would have been deduced from SIDM-only simulations. Our results imply that in SIDM models the distributions of dark matter and baryons in galaxies are strongly correlated. 

{\em Acknowledgements:}  We thank James Bullock and Sean Tulin for useful discussions. MK is supported by NSF Grant No. PHY-1214648. TL is supported by the National Aeronautics and Space Administration through Einstein Postdoctoral Fellowship Award Number PF3-140110. HBY is supported by startup funds from UCR. 

\bibliography{sidmcore}

\end{document}